\documentclass[aps,pre,superscriptaddress,groupaddress,twocolumn]{revtex4}

\usepackage[justification=RaggedRight,labelfont=bf,format=plain,font=small]{caption}
\usepackage{changepage}
\usepackage{amsmath}    
\usepackage{graphicx}   
\usepackage{verbatim}   
\usepackage{color}      
\usepackage{subfigure}  
\usepackage{xr}
\begin{document}
\makeatletter 

\renewcommand{\figurename}{{\bf Figure}}
\makeatother

\title{A minimal model for spontaneous cell polarization and edge activity in oscillating, rotating and migrating cells}

\author{Franck Raynaud}
\affiliation{Laboratory of Cell Biophysics, Ecole Polytechnique F\'ed\'erale de Lausanne, 1015 Lausanne, Switzerland}
\author{Mark E. Amb\"uhl \footnote{Now at SICHH, Swiss Integrative Center for Human Health, Fribourg, Switzerland}}
\affiliation{Laboratory of Cell Biophysics, Ecole Polytechnique F\'ed\'erale de Lausanne, 1015 Lausanne, Switzerland}
\author{Chiara Gabella}
\affiliation{Laboratory of Cell Biophysics, Ecole Polytechnique F\'ed\'erale de Lausanne, 1015 Lausanne, Switzerland}
\author{Alicia Bornert}
\affiliation{Laboratory of Cell Biophysics, Ecole Polytechnique F\'ed\'erale de Lausanne, 1015 Lausanne, Switzerland}
\author{Ivo F. Sbalzarini}
\affiliation{MOSAIC Group, Chair of Scientific Computing for Systems Biology, Faculty of Computer Science, TU Dresden, D-01069 Dresden, Germany}
\affiliation{MOSAIC Group, Center for Systems Biology Dresden, Max Planck Institute of Molecular Cell Biology and Genetics, D-01307 Dresden Germany}
\author{Jean-Jacques Meister}
\affiliation{Laboratory of Cell Biophysics, Ecole Polytechnique F\'ed\'erale de Lausanne, 1015 Lausanne, Switzerland}
\author{Alexander B. Verkhovsky}
\affiliation{Laboratory of Cell Biophysics, Ecole Polytechnique F\'ed\'erale de Lausanne, 1015 Lausanne, Switzerland}

\begin{abstract}
How the cells break symmetry and organize their edge activity to move directionally is a fundamental question in cell biology. Physical models of cell motility commonly rely on gradients of regulatory factors and/or feedback from the motion itself to describe polarization of edge activity. Theses approaches, however, fail to explain cell behavior prior to the onset of polarization. Our analysis using the model system of polarizing and moving fish epidermal keratocytes suggests a novel and simple principle of self-organization of cell activity in which local cell-edge dynamics depends on the distance from the cell center, but not on the orientation with respect to the front-back axis. We validate this principle with a stochastic model that faithfully reproduces a range of cell-migration behaviors. Our findings indicate that spontaneous polarization, persistent motion, and cell shape are emergent properties of the local cell-edge dynamics controlled by the distance from the cell center.
\end{abstract}

\maketitle

The ability to break symmetry and move directionally is an essential property of most eukaryotic cells~\cite{Ridley2003,Danuser2013,Mogilner2009}. This happens in response to external stimuli, but also spontaneously~\cite{Wedlich2003,Devreotes2003,Verkho1999}. Persistent motion requires segregation of cell-edge activities, so that protrusion happens predominantly at the front, and retraction at the back of the cell. In contrast, in cells exploring their environment, edge activity is on average spatially isotropic, but fluctuates in time between protrusion and retraction~\cite{Tkachenko2011,Burnette2011,Ryan2012,Giannone2007,Ji2008}. Thus both exploratory activity and the directional motion depend on the transitions between protrusion and retraction but how the cell chooses between these two regimes to establish spatial and temporal patterns of edge activity remains unclear. It is believed that in migrating cells a directional mechanism at the scale of the whole cell, e.g. a global gradient of cytoskeletal and/or signaling components, orchestrates cell-edge dynamics according to the overall motion direction~\cite{Ridley2003,Lakayo2007,Devreotes2003,Lin2012,Keren2008} This concept is limited in that external directional stimuli~\cite{Devreotes2003,Lin2012} in combination with internal diffusible signals interacting through feedback loops~\cite{Onsum2009,Holmes2012,Maree2006,Mori2008,Satulovsky2008,Doubrovinski2011}, feedback from the motion itself~\cite{Verkho1999,Nishimura2009,Ziebert2012,Du2012,Maiuri2015}, or very large and highly correlated perturbation to induce the polarized states~\cite{Barnhart2015} have to be invoked in order to establish the polarity axis.\\
Fish epidermal keratocytes, thanks to their robust polarity, simple shape, and persistent motion, are a classic model system to study polarization and directional migration~\cite{Mogilner2009,Lee1993}. Analysis of the edge dynamics of these cells led to the concept of Graded Radial Extension (GRE), which links local edge dynamics to the resulting overall cell motion: protrusion and retraction are directed normally to the cell edge with rates that are graded depending on the orientation with respect to the motion direction~\cite{Lee1993}. This model inspired several studies searching for the underlying mechanism in the form of a gradient of cytoskeletal and/or signaling components from the front to the back of the cell. Gradients of myosin II, actin, and small GTPases of the Rho family~\cite{Lakayo2007,Barnhart2011,Maree2006,Verkho1999,Ziebert2012,Grimm2003,Keren2008,Wilson2010} have been implicated, but it remains unclear how these gradients arise in the first place and whether they are cause or consequence of polarization and directed motion~\cite{Wolgemuth2011}. The concept of a directional gradient guiding the cell polarity axis cannot describe cell-edge activity in presence of multiple leading edges~\cite{Lou2015} or prior to the point when the direction of motion becomes defined. In this study, we start from quantification of experimentally observable cell edge dynamics to identify features that are common to the migration and the polarization process. We find that cell edge universally displays a distance-dependent switch from protrusion to retraction, and we develop a computational model to demonstrate that this property is sufficient for spontaneous polarization and directional motion.

\section{Edge dynamics in migrating cells}
\begin{figure*}
\includegraphics[clip=true,width=0.75\textwidth]{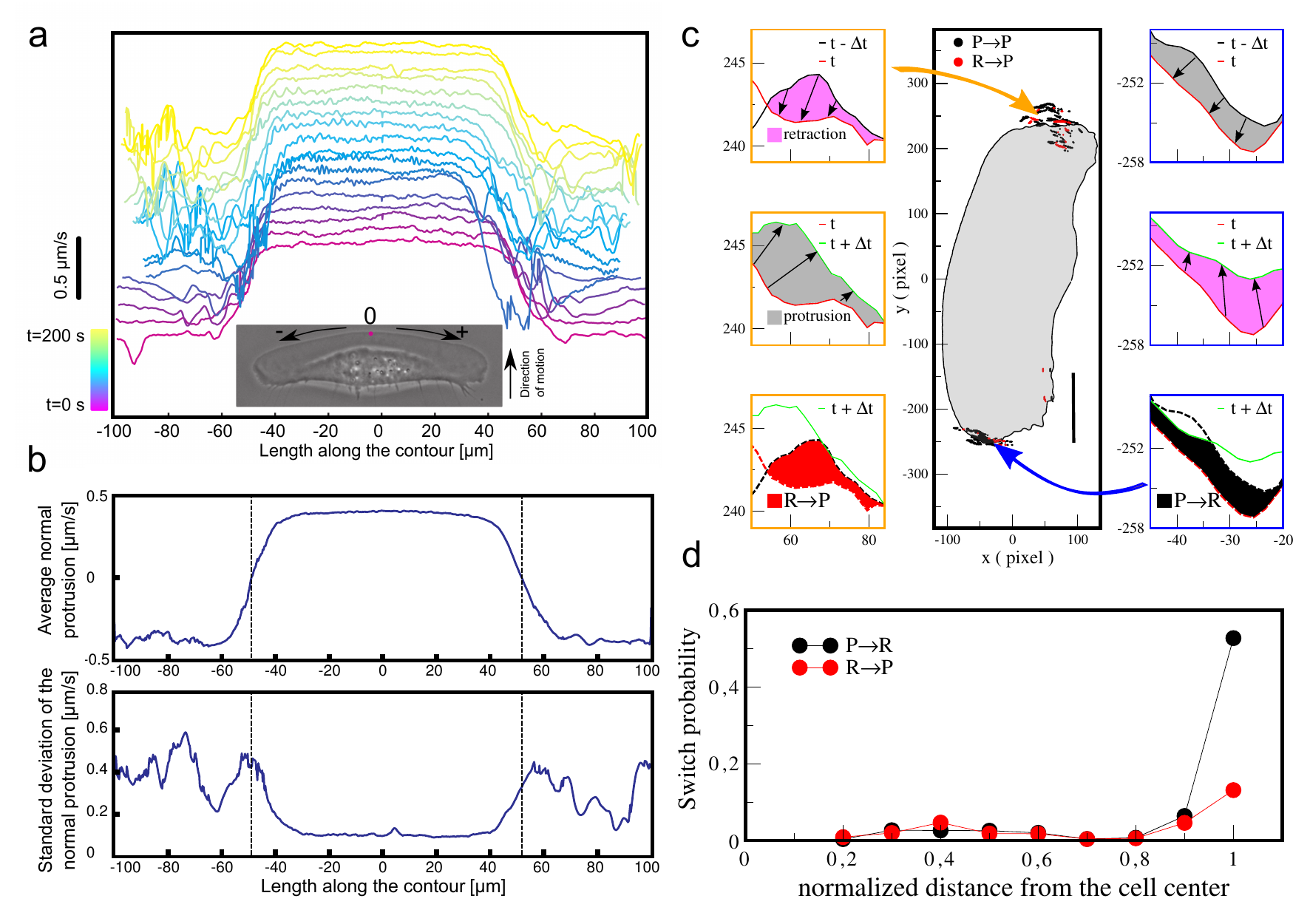}
\caption{{\bf Edge dynamics and distribution of switches between protrusion and retraction in migrating cells.}
(a) Normal edge velocity profiles as function of position along the cell contour averaged over consecutive 11 s intervals. Time is color-coded from purple to yellow and profiles are spaced by 0.2$\mu$m/s for clarity. The zero contour position (middle of the leading edge) is illustrated on the cell image (inset). (b) Mean normal velocity along the cell contour (top) and standard deviation of the velocity (bottom) over 199s. Vertical lines mark the positions of zero mean velocity. (c) Localization of PR (black) and RP (red) switches during 82s of migration plotted in the co-moving frame.(central panel). Detection of switching regions as intersections of protrusion (gray) and retraction (pink) in three consecutive contours (SI) (left and right panels). (d) Switch probability as function of the normalized distance from the cell center. Distances were normalized by the maximum distance from the cell center in order to aggregate data from cells of different size. Contours were extracted every 2s from five different cells for a total of 236 contours. Scale bar is 6.45$\mu$m\label{fig1}}
\end{figure*}
Using a recently developed cell-edge segmentation and tracking method~\cite{Markus2012}, we have quantified protrusion velocities in keratocytes with high spatial and temporal resolution. During migration, the protrusion velocity in the regularly shaped (coherent) cells~\cite{Lakayo2007,Keren2008} was nearly constant and stable in time along the central part of the leading edge, but declined and fluctuated toward the sides, becoming eventually negative (retraction) (Fig.~\ref{fig1}a and b). While previous studies considered the edge velocity to be constant in time and rather emphasized its gradation in space, we explore the outcome of temporal velocity fluctuations. We designed a method to identify the positions along the cell contour where the edge velocity switched in time from protrusion to retraction (PR) and from retraction to protrusion (RP) (Fig.~\ref{fig1}c and Methods). Our mapping revealed that most of the switches localized to narrow zones at the lateral extremities of the cells. Plotting the switch probability as a function of the distance from the geometrical center of the cell (see Methods) indicated that approximately $70\%$ of switches occurred between $90$ and $100\%$ of the maximal cell extension (Fig.~\ref{fig1}d). The majority of switches were from protrusion to retraction, but, interestingly, we observed that the cell extremities also displayed RP switches ($\sim15\%$ the total number of switches), suggesting metastable edge dynamics at these sites. Some switches marginally appeared ($\le 5\%$) at intermediate distances (Fig.~\ref{fig1}d). They occurred at the back of the cell and were attributed to the dynamics of retraction fibers~\cite{Schaub2007}.\\ 
\begin{figure}
\includegraphics[width=0.5\textwidth]{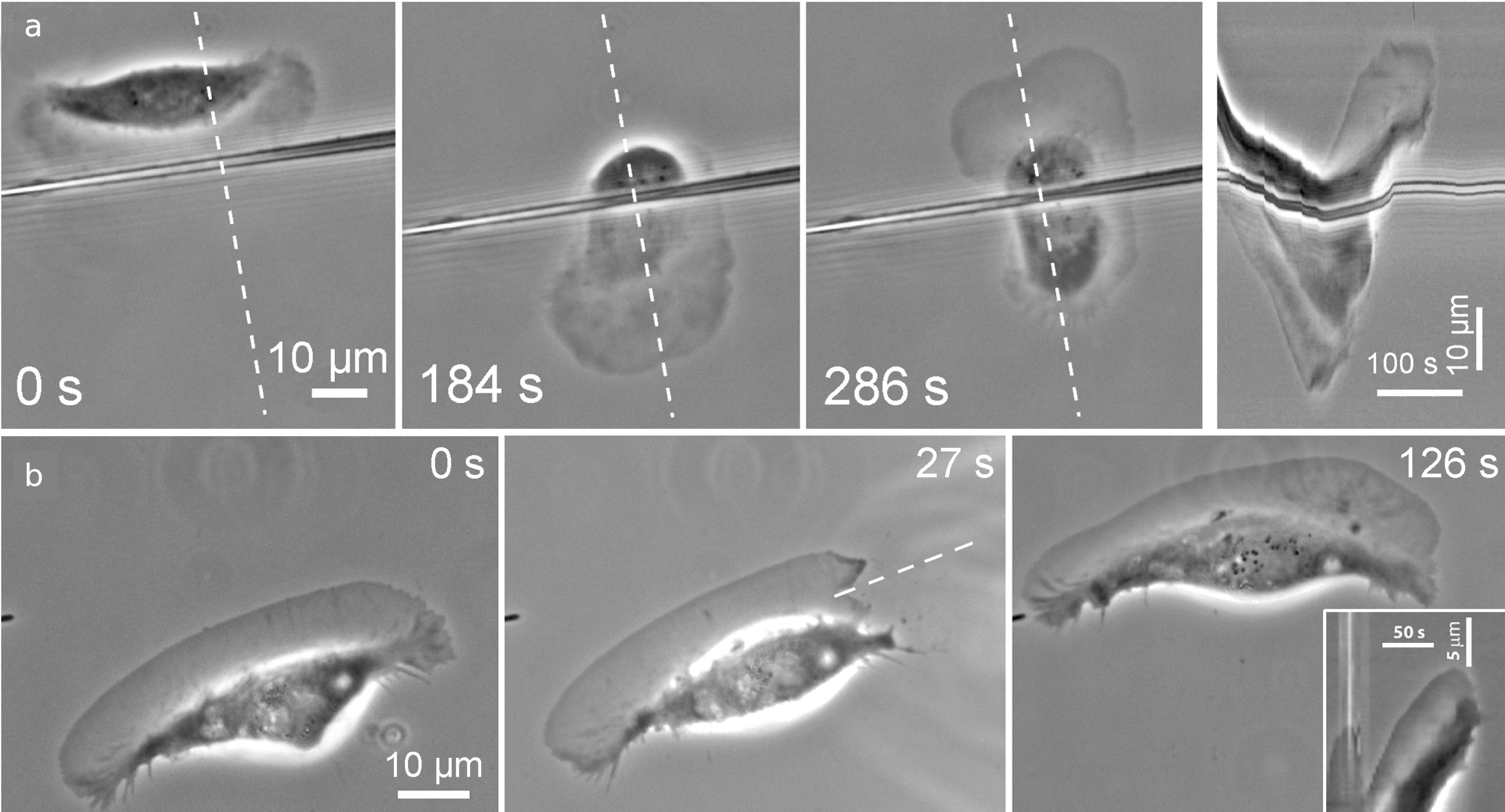}
\caption{{\bf Protrusion-retraction switches depend on the distance from the cell center, but not on the orientation.} 
(a) Snapshots (left) and kymograph (right) of the cell with its body stalled by a pipette. Time is indicated in seconds after encounter with the pipette. (b) Snapshots of the cell before, 1s after, and 100s after its right lateral part was cut off with a glass needle. Persistent protrusion at the cut site results in shape recovery and turning in the direction of cut. Inset in the right panel is kymograph taken along the white dashed line shown in the middle panel. Kymographs were generated from the sequences taken at 2s (a) and 3s (b) frame intervals.\label{fig2}}
\end{figure}

\begin{figure*}
\includegraphics[width=0.9\textwidth]{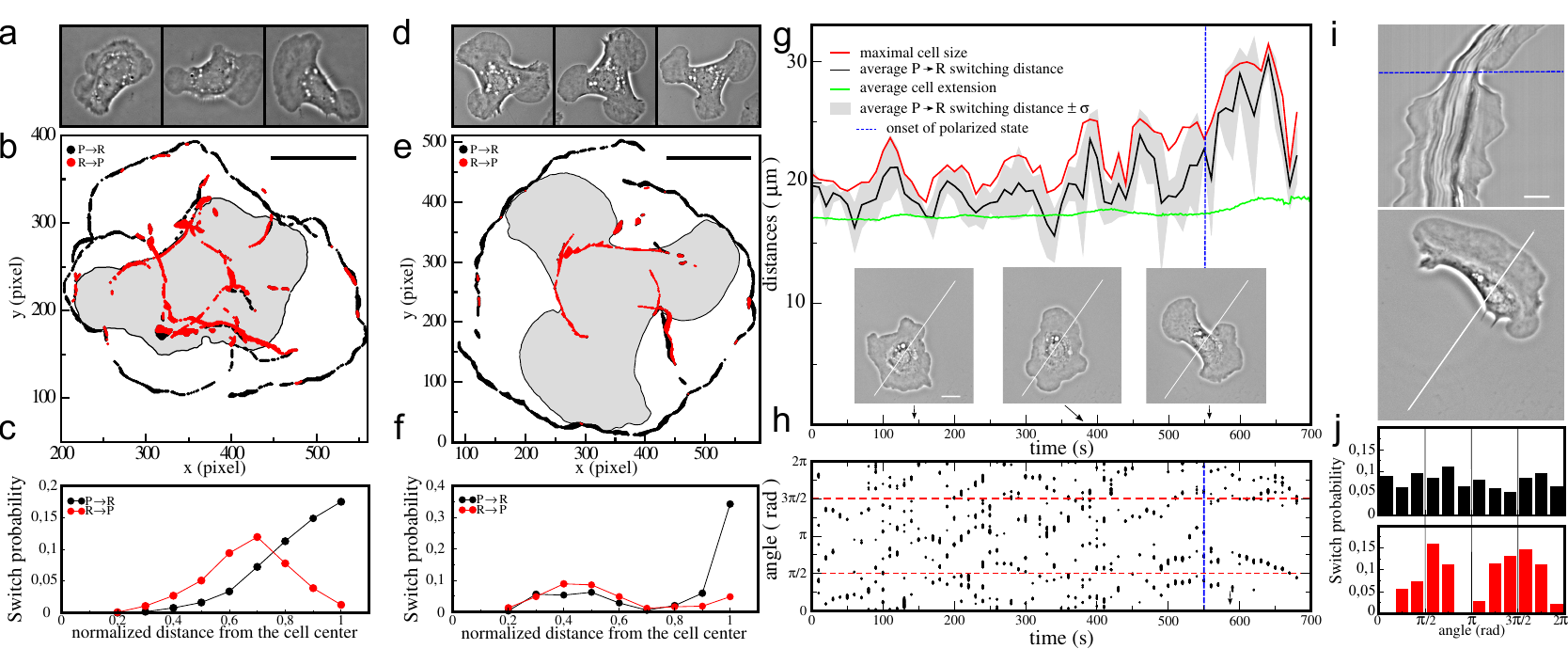}
\caption{{\bf Distribution of switches between protrusion and retraction and shape properties during polarization.}
Phase-contrast snapshots of fluctuating (a) and rotating (d) cells. Localization of PR (black) and RP (red) switches during 366s of fluctuation (b) and 304s of rotation (e).  (c) and (f) Switch probability as function of the normalized distance from the cell center in polarizing (c) and rotating (f) cells. Contours of polarizing cells were extracted every 10s (5 cells, 214 contours in total). Contours of rotating cells were extracted from two cells in 2-fold and 3-fold rotation every 6s and 4s, respectively, for a total number of 110 contours. Scale bar is  6.45$\mu$m (polarizing) and  10.75$\mu$m (rotating). (g) Time course of maximal cell extension (red) average cell extension (green) and average PR switching distance (black) during the process of polarization. Snapshots of the cell at indicated times during polarization are shown in the insets. (i) Kymograph (upper panel) of the polarizing cell taken along the white line shown in the insets in (g). Transition time point (t=550s) is indicated in the kymograph (blue dashed line) when edge fluctuations stop and the cell starts moving. At this point the cell is not yet well polarized and the shape remains irregular (right inset in (g)). At the end of the sequence the cell is fully polarized (lower panel). In (g) and (i) the scale bar is 10$\mu m$. (h) Map of the time course of the angular distribution of protrusion-retraction switches (black dots). The angles are taken with respect to the direction of the instantaneous cell velocity. (j) Protrusion-retraction switch probability as function of the angle before (upper panel) and after the transition point (lower panel).\label{fig3}}

\end{figure*}
We point out that the regions of most probable switching correspond at the same time to the maximal cell extension and to the parts of the cell edge displacing orthogonally to the direction of motion. To discriminate which of the two descriptors, orthogonal orientation with respect to the direction of motion or distance from the cell center, defines the distribution of switches, we manipulated the cells to change the distance between the edge and the cell center at different orientations. To increase the distance along the direction of motion, we placed a micropipette in front of the cell, allowing the leading edge to squeeze between the pipette and the substrate, but blocking the taller nucleus. The leading edge maintained its directional motion at constant velocity (right panel in Fig.~\ref{fig2}a) until the cell extended to a distance similar to its original maximal lateral extension. At this point, the edge halted, fluctuated, and switched to retraction, resulting in a reversal of cell motion (Fig.~\ref{fig2}a and Supplementary Video~\ref{vidpipette}, similar responses observed in 12 out of 12 cells). We also tested the effect of a reduction of the distance between the edge and the cell center by cutting off a part of one lateral wing of a cell. The cut side exhibited persistent protrusion transverse to the direction of motion (inset in Fig.~\ref{fig2}b) until the cell recovered its original width. The protrusion zone may even propagate from the site of the cut to the back of the cell, resulting in a turning of the cell in the direction of the cut (Fig.~\ref{fig2}b and Supplementary Video~\ref{vidcut}, identical response observed in 29 out of 32 cells). Since pipette blocking induced switches at the front and cutting suppressed them at the side, both types of manipulations provide evidence that switches are regulated by distance from the cell center rather than by angle with respect to the cell motion direction. If angular hypothesis was true, the cell would continue protruding at the front in the pipette experiment and maintained switches at the cut side in the cutting experiment.

\section{Edge dynamics in polarizing cells}

Next, we investigated the dynamics of edge activity during the process of spontaneous cell polarization. In the course of polarization, most of the cells exhibited multiple irregular fluctuations and waves around the edge with small regions of protrusion and retraction coalescing into larger zones~\cite{Markus2012} (Fig.~\ref{fig3}a, Supplementary Videos~\ref{vidalicia} and \ref{vidrotpol}). Polarized cell with a single protruding and a single retracting region emerged as the result of these protrusion-retraction fluctuations that occurred during the polarization process. Interestingly some cells were trapped during polarization in exotic 2-fold or 3-fold symmetric configurations with several leading edges rotating for several minutes (Fig.~\ref{fig3}d and Supplementary Video~\ref{vidrotpol}). Both fluctuating and rotating cells eventually polarized forming a single protruding and a single retracting region. Mapping of switches in fluctuating cells showed that switch sites were distributed evenly along the cell edge  (Fig.~\ref{fig3}b) and both types of transition had similar probability to occur ($55\%$ of the total number were PR switches). The distribution of switching distances displayed a peak of PR switches near the maximal cell extension and a peak of RP switches at intermediate distances (Fig.~\ref{fig3}c). In the cells that displayed multiple rotating edges, switches localized mostly at the tips and bases of the rotating segments (Fig.~\ref{fig3}e), and the two types of switches were unbalanced ($65\%$ of PR switches). As in fluctuating and migrating cells, PR switches occurred near the maximal cell extension, while RP switches were more likely at short distances from the cell center (Fig.~\ref{fig3}f).\\

Analysis of dynamics of PR switches in time during the transition between isotropic and polarized states demonstrated that they remained near the maximal cell extension during the entire process of polarization, including the onset of persistent migration (Fig.~\ref{fig3}g). Measuring the angle between the orientation of PR switches with respect to the cell center and the direction of the instantaneous velocity showed that switches occurred at any angle up to the onset of polarized state and then the angles became narrowed to the two zones orthogonal to the velocity (Fig.~\ref{fig3}h and j). These results indicate that despite changes in cell morphology the switches from protrusion to retraction always occur near maximal cell extension  suggesting that this is an important feature underlying cell-edge dynamics during the polarization process.\\

\begin{figure}[t!]
\includegraphics[width=0.5\textwidth]{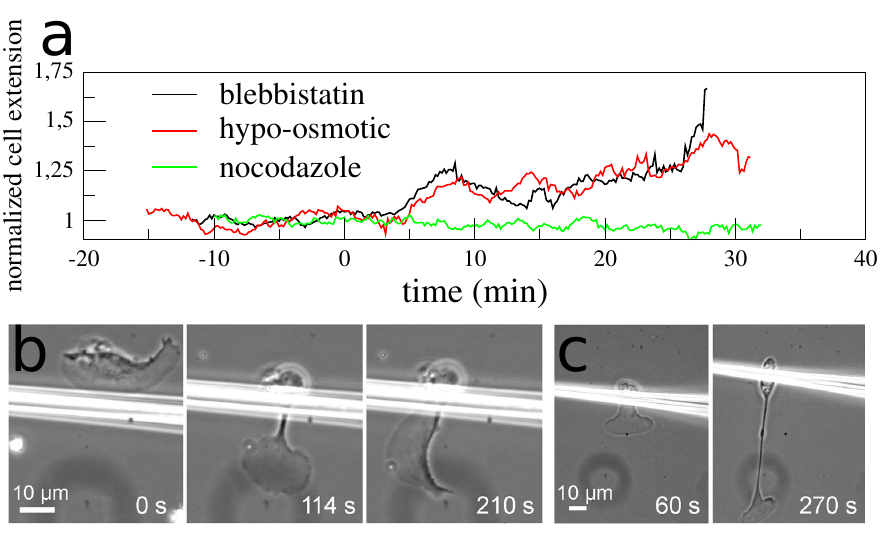}
\caption{{\bf Effect of blebbistatin, hypo-osmotic treatment, and nocodazole on cell width and behavior in body-blocking experiments.} (a) Dynamics of the average normalized cell width after treatment with blebbistatin (black, 11 cells), nocodazole (green, 10 cells), and hypo-osmotic shock (red, 9 cells). The width of each cell was normalized with its average before treatment. (b) A blebbistatin-treated cell extends its front on a narrow stalk after body blocking and then turns parallel to the pipette; (c) the cell after hypo-osmotic shock keeps extending its front.\label{fig4}}
\end{figure}

In order to gain insight into the mechanisms controlling this switching distance we tested if the cell width during migration and the response in body-blocking experiments depended on the cell volume, contractility, and the integrity of the microtubule system. Increasing the cell volume by hypo-osmotic treatment and inhibiting myosin-II-dependent contractility by blebbistatin both led to an increase in cell width (with eventual cell fragmentation in the case of myosin inhibition)  (Fig.~\ref{fig4}a). Cell-body-blocking experiments with inflated cells showed a partial suppression of the distance sensing: in $50\%$ of the cases (3 out of 6 cells) cells kept extending their leading edge until it eventually detached from the cell body (Fig.~\ref{fig4}c and Supplementary Video~\ref{vidpipette}). The blebbistatin-treated cells extended their front lamellipodia away from the cell body, while remaining connected to the latter by a narrow stalk (in contrast to a wide bridge in the control case). However, in $75\%$ of the cases (9 out of 12 cells) the lamellipodia did not extend continuously, instead turning to a motion direction parallel to the pipette (Fig.~\ref{fig4}b and Supplementary Video~\ref{vidpipette}). These results indicate that distance sensing is not a trivial consequence of finite membrane area or of cytoskeletal protein content, but depends on the cell volume and/or three-dimensional shape~\cite{Chiara2013} along with myosin-dependent contractility. In contrast, disassembly of microtubules with nocodazole affected neither the cell width (Fig.~\ref{fig4}a), nor the cell response in the body-blocking experiments (Supplementary Video~\ref{pipettenoco}). This suggests that, unlike in other systems where micro-tubules were implicated in cell-size control~\cite{Martin2009,Picone2010}, they do not play such a role in keratocytes.

\section{Computational model}
\begin{figure*}
\includegraphics[clip=true,width=0.9\textwidth]{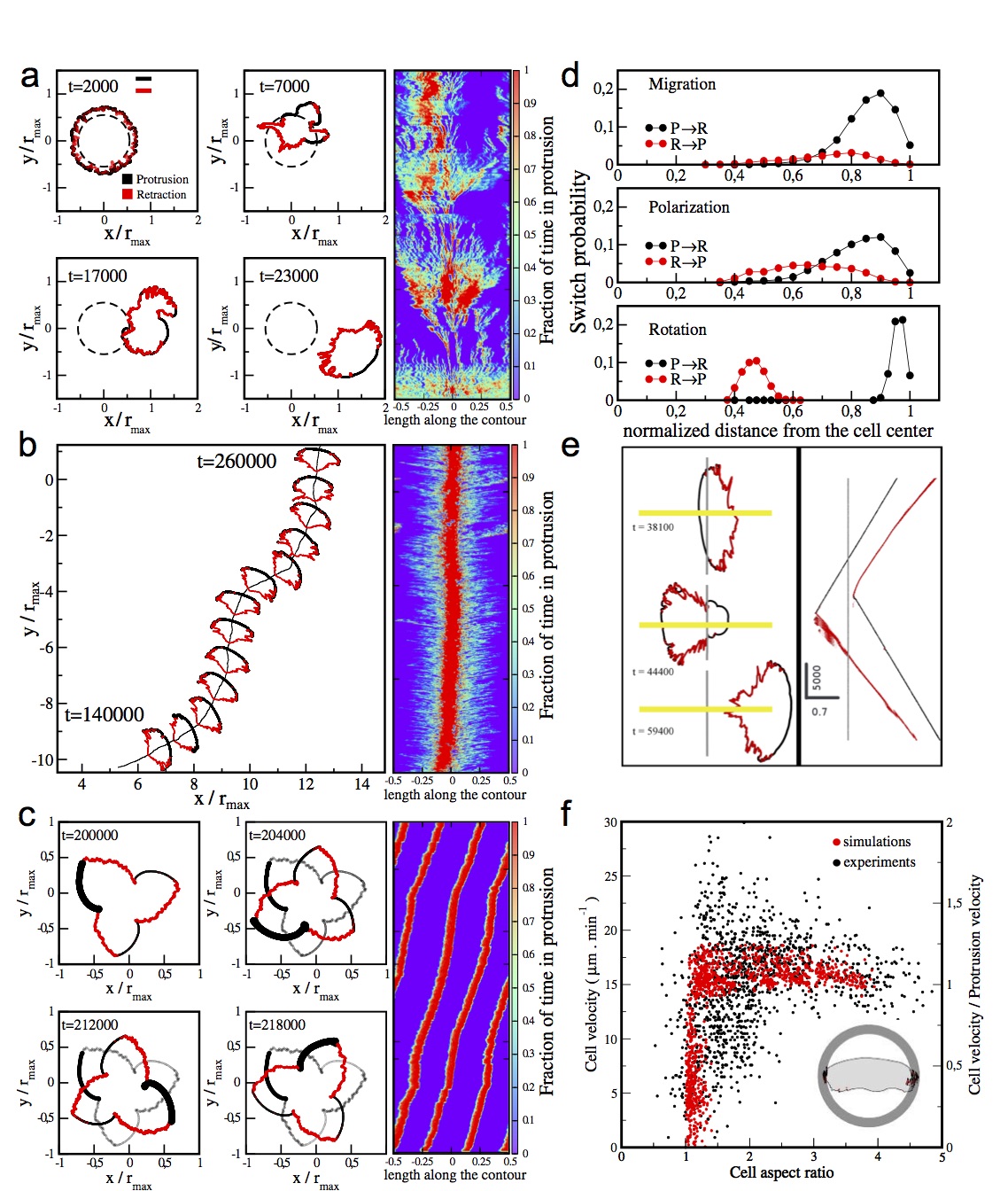}
\caption{{\bf Computer model simulations of cell-edge dynamics.}
Successive edge configurations during polarization (a), migration  (b, snapshot every 10,000 time steps ) and rotation (c) with the corresponding maps of protrusion (right panels). Protruding (retracting) points are shown in black (red), and the axes are in units of rmax. In the protrusion maps, the fraction of time spent in protrusion (color scale) is computed for each point in a sliding window of 100 time steps. Zero curvilinear abscissa corresponds to the intersection of the cell edge with the line emerging from the cell center in the average direction of motion. Positive (negative) coordinates stand for clockwise (counterclockwise) directions along the contour. The simulations were run with a range of interaction $N=4$ and a noise $\sigma=0.15 r^{max}$. The points had random initial states and were initially equi-distributed on a circle. The simulation for the rotation was run with the following parameters: $V^{max} = 1.2 V_p, V^{min} = 0.8 V_p$, $N = 48$, and the coupling term $\left(\frac{n}{N}\right)^5$. (d) Switch probability as function of the normalized distance from the cell center computed from the sequences shown in (b) and (c) over 10 independently generated simulations of polarization. Contours were recorded every 500 time steps. (e) Numerical experiment with a polarized cell facing an obstacle (grey line). Left panel: the cell is in contact with the obstacle (top), then it is blocked (middle), and finally it backs up from the obstacle and regains its polarized shape in the opposite direction (bottom). Right panel: kymograph extracted from the sequences along the yellow line. Scale bars are shown in number of time steps (vertical) and units of rmax (horizontal). (f) Instantaneous aspect ratio and velocity during migration for experiments (black) and simulations (red).  Experimental points were obtained from 21 cells (total migration time 15 h, frame rate 1/30 s). To simulate cells with different aspect ratios, we modify the threshold distance rmax while keeping all other parameters constant. Simulations lasted for $10^6$ time steps for each rmax; velocity and aspect ratio are reported every 5,000 time steps.\label{fig5}}
\end{figure*}
We developed a stochastic model incorporating the features of cell-edge dynamics uncovered in our experiments. The model is deliberately reductionist and minimal, in order to test whether switches induced by a threshold distance are indeed sufficient to explain the transition to a polarized state, directional motion and cell shape. We describe the cell edge by a set of points (i.e., computational particles) ${\vec{X}_i}$ in two possible states: protrusion (P) or retraction (R). PR switches occur at a threshold distance $r_{max}$ from the cell center, while RP switches occur at a smaller threshold distance  $r_{min}$.  The state of a point also changes if most of its neighbors within a defined interaction range are in the opposite state, providing coupling that enables local ordering of the edge activity. Specifically, each point $i$, at a distance $r$ from the cell center has a probability $P_{P\rightarrow R}$ (resp. $P_{R\rightarrow P}$) to switch from protrusion to retraction (resp. from retraction to protrusion), given by:

\begin{equation*}
P_{P\rightarrow R}=
\begin{cases}
   \tau \biggl(\mathcal{N}_{r_{max},\sigma^2} +  \left(1-\mathcal{N}_{r_{max},\sigma^2}\right)\frac{n_i}{N}\biggr) & \text{if } r<r_{max},\\
\tau & \text{if } r>r_{max},
\end{cases}
\end{equation*}
\begin{equation*}
P_{R\rightarrow P}=
\begin{cases}
  \tau\frac{n_i}{N} & \text{if } r>r_{min},\\
\tau & \text{if } r<r_{min},
\end{cases}
\vspace*{0.05in}
\end{equation*}

where $\tau$ is an overall rate of transition ($\Delta t^{-1}$) that takes into account the self-persistence of the point state. It effectively allows increasing the displacement of a point between transitions while preventing numerical instabilities. $\mathcal{N}_{r_{max},\sigma^2}$ is a Gaussian random number with mean  $r_{max}$ and variance $\sigma^2$ (corresponding to a noisy threshold distance), $n_i$ is the number of neighbors (within the range of interaction N) of $i$ that are in the opposite state, $\frac{n_i}{N}$ thus accounts for coupling. The position of each point $i$ is then updated as:\\

\begin{equation*}
\vec{X}_i\left(t+\Delta t\right) = \vec{X}_i\left(t\right) + \vec{V}_{p,r}\Delta t,
\vspace*{0.05in}
\end{equation*}

with protruding nodes moving outward normally to the edge at constant velocity $V_p$ (Fig.~\ref{fig1}a) and  retracting nodes moving toward the cell center $\left(x_c, y_c\right)$ with a distance dependent velocity $V_r(r)$ (Methods and Supplementary Fig.~\ref{figS2}). At each time step, the position of the cell center is calculated from the coordinates of edge points $\left(x_i, y_i\right)$ as described in Methods.

For a wide range of parameters (Supplementary Figs.~\ref{figS3},~\ref{figS4},~\ref{figS5}), this combination of isotropic rules of local edge dynamics led to self-organization with spontaneous transition from disordered initial states to a polarized motile state with emergent shapes reproducing experimental observations (Fig.~\ref{fig5}). The simulations were started from a circular outline with random distribution of protruding and retracting points; first, the edge displayed strong shape fluctuations comparable to those observed in experiments with small protruding and retracting regions traveling and fusing into larger zones (Fig.~\ref{fig5}a, Supplementary Video~\ref{simulpolarization}). Then, the persistently migrating state was reached with stable typical keratocyte shapes elongated perpendicular to the direction of motion (Fig.~\ref{fig5}b, Supplementary Video~\ref{simulpolarization}). The analysis of the resulting cell outlines with the same switch mapping protocol that was used for real cells yielded distributions of switches similar to the experimental ones both for polarizing and migrating cells (Fig.~\ref{fig5}d). The model was also able to sustain the rotating state of a specific initial 3-fold configuration with three leading and trailing edges (Fig.~\ref{fig5}c and Supplementary Video~\ref{simulrot}). We also simulated cell-blocking experiments with a virtual pipette in the form of a line that could not be passed by the cell center. In this numerical experiment, the outline backed up in a way similar to real cells: by extending at a constant speed, halting, fluctuating in place, and eventually retracting and reversing its direction (Fig.~\ref{fig5}e, and Supplementary Video~\ref{simulpipette}). Finally, removing the lateral part of the cell outline and replacing it with a straight line segment resulted in persistent protrusion and turning of the cell toward the removed part of the edge (Supplementary Video~\ref{simulcut}), in agreement with the cutting experiments.\\
The simulation model also allowed us to test different mechanisms of shape feedback. If the distance limit was replaced by linking the switching probability to either cell area or to the overall ratio of protruding vs. retracting points, the system developed neither stable shape nor directional motion (Supplementary Video~\ref{simulpolmig}). Any feedback coupling local edge dynamics to a mean cell property that affects all points would prevent local order from propagating. In contrast, the distance limit, while still encoding global information through the position of the cell center, affects only specific points. A distance threshold coupled to the motion of the cell creates localized zones of frustration at the cell edge, where PR switch is favored, maintaining phase segregation of cell-edge activity and hence stabilizing the system. Experimentally, only the lateral extremities of migrating cells reach the critical distance and switch from protrusion to retraction, thus stabilizing the width of the cell, while the points at the front and back move in concert and maintain their state as long as they do not pass the critical distance threshold. This switch distribution pattern is only possible if the cell has an elongated shape, explaining why such anisotropic shapes, minimizing the edge fluctuations, emerge from the isotropic distance-threshold mechanism. This is consistent with keratocytes being one of the most efficient types of migrating cells. If, for any reason, the aspect ratio of the cell decreases and approaches unity, our model predicts that the distance threshold would induce switches to retraction at the front and fluctuations of the velocity of the leading edge, resulting in a decrease of the net velocity of motion. Indeed, it has been experimentally observed that keratocyte velocity depends on the cell aspect ratio with cells of smaller aspect ratio having variable and lower velocities~\cite{Keren2008}. We have confirmed this effect experimentally and also reproduced it in the simulations (Fig.~\ref{fig5}f).

Protrusion/retraction transition is a ubiquitous cellular phenomenon that has been linked with various molecular processes~\cite{Tkachenko2011,Burnette2011,Ryan2012,Giannone2007} and with mechanical force generation~\cite{Ji2008}. However, previous studies did not provide a unified model to relate local cell-edge fluctuations to either cell symmetry breaking during polarization or overall cell shape and motion. In this paper, we used high-resolution edge segmentation and cell-shape modification experiments, to show that the dynamics of switches from protrusion to retraction is controlled by the distance from the cell center. We developed a top-down stochastic model to demonstrate that this isotropic property of distance-dependence leads to spontaneous symmetry breaking. Our model thus relates overall cell behavior to local protrusion-retraction dynamics. The model simulations have shown that the hypothesized mechanism is indeed sufficient to reproduce the experimentally observed cell behavior, also under perturbations. Such distance sensing leads to a genuine self-organization with emerging shape and motion being the macroscopic manifestations of rotational symmetry breaking in the localization of switches. Our findings suggest that cell volume and contractility, but not the microtubule system, may play a role in the mechanism of distance sensing. One possibility related to contractile properties is that traction stresses generated by acto-myosin network scale with the size of the network and induce detachment from the substrate and collapse at a critical distance. Another possibility is that the distance and volume can influence force balance at the cell edge through three-dimensional geometry: extension of the cell is expected to flatten its apical surface and therefore increase the components of membrane~\cite{Chiara2013} and cortical tension~\cite{Fouchard2014} in the substrate plane, while increasing the cell volume would have the opposite effect on the orientation of forces and thus suppress distance sensitivity. These force-related mechanisms do not exclude a contribution of intra-cellular gradients of structural or regulatory components, which may depend on cell volume and contractility. A unique feature of our model, however, is that gradients do not necessarily have to be oriented in the direction of motion. Instead, we propose a center-to-periphery radial gradient of the probability of protrusion-retraction switches. This probability may depend on regulatory factors, e.g. small GTPases, that could develop a radial gradient by a reaction-diffusion mechanism~\cite{Holmes2012,Mori2008}, marking the cell margins in a way that is similar to how the Min-protein system in bacteria marks the middle of the cell~\cite{Schweizer2012}. After motion onset, front-to-back gradients may develop as well, but their role would be to reinforce motion rather than to initiate it. Another important feature of our model is that cell-edge activity is not imposed deterministically. Instead, the position in the putative radial gradient only encodes disjoint zones of increased switching probability. The behavior between these zones may rely on cytoskeletal reactions favoring persistence, e.g., autocatalytic actin network branching~\cite{Pollard2007} and contraction that may be self-sustained due to its ability to concentrate myosin motors~\cite{Verkho1999,Barnhart2015,Schaub2007}. Future studies will uncover the interplay of cell contractility, cell volume, and other factors involved in radial organization in order to reveal the physical and molecular underpinnings of distance sensing and to determine how this mechanism is involved in different cell shapes and behaviors.\\

\section{Acknowledgements}
 We would like to thank F. N\'ed\'elec and M. Balland for useful discussions, S. Bohnet for the first observation of cell rotation, and H. Troyon for experimental assistance. This work is supported by Swiss National Science Foundation Grant 31003A-135661. M.E.A. was funded by a PhD fellowship from the Swiss National Competence Center for Biomedical Imaging, to A.V. and I.F.S. 
\section{Author contributions}
F.R. and A.B.V. designed the study, M.E.A., C.G., A.B. and A.B.V. performed the experiments, F.R. and M.E.A. analysed the data, F.R. developed the numerical model, F.R., I.F.S. and A.B.V. wrote the paper. All authors have discussed the results and the interpretation. F.R. and M.E.A. have contributed equally to the study.

\end{document}


\date{}
\title{A minimal model for spontaneous cell polarization and edge activity in oscillating, rotating and migrating cells}

\author[1]{Franck Raynaud}
\author[1]{Mark E. Amb\"uhl\footnote{Now at SICHH, Swiss Integrative Center for Human Health, Fribourg, Switzerland}}
\author[1]{Chiara Gabella}
\author[1]{Alicia Bornert}
\author[2,3]{Ivo F. Sbalzarini}
\author[1]{Jean-Jacques Meister}
\author[1]{Alexander B. Verkhovsky}
\affil[1]{Laboratory of Cell Biophysics, Ecole Polytechnique F\'ed\'erale de Lausanne, 1015 Lausanne, Switzerland}
\affil[2]{MOSAIC Group, Chair of Scientific Computing for Systems Biology, Faculty of Computer Science, TU Dresden, D-01069 Dresden, Germany}
\affil[3]{MOSAIC Group, Center for Systems Biology Dresden, Max Planck Institute of Molecular Cell Biology and Genetics, D-01307 Dresden Germany}

\maketitle
\newpage
\section*{{\Large Supplementary Informations}}
\makeatletter 
\renewcommand{\figurename}{{\bf Supplementary Figure}}

\begin{figure*}[h!]
\centering \includegraphics[ clip=true,width=\textwidth]{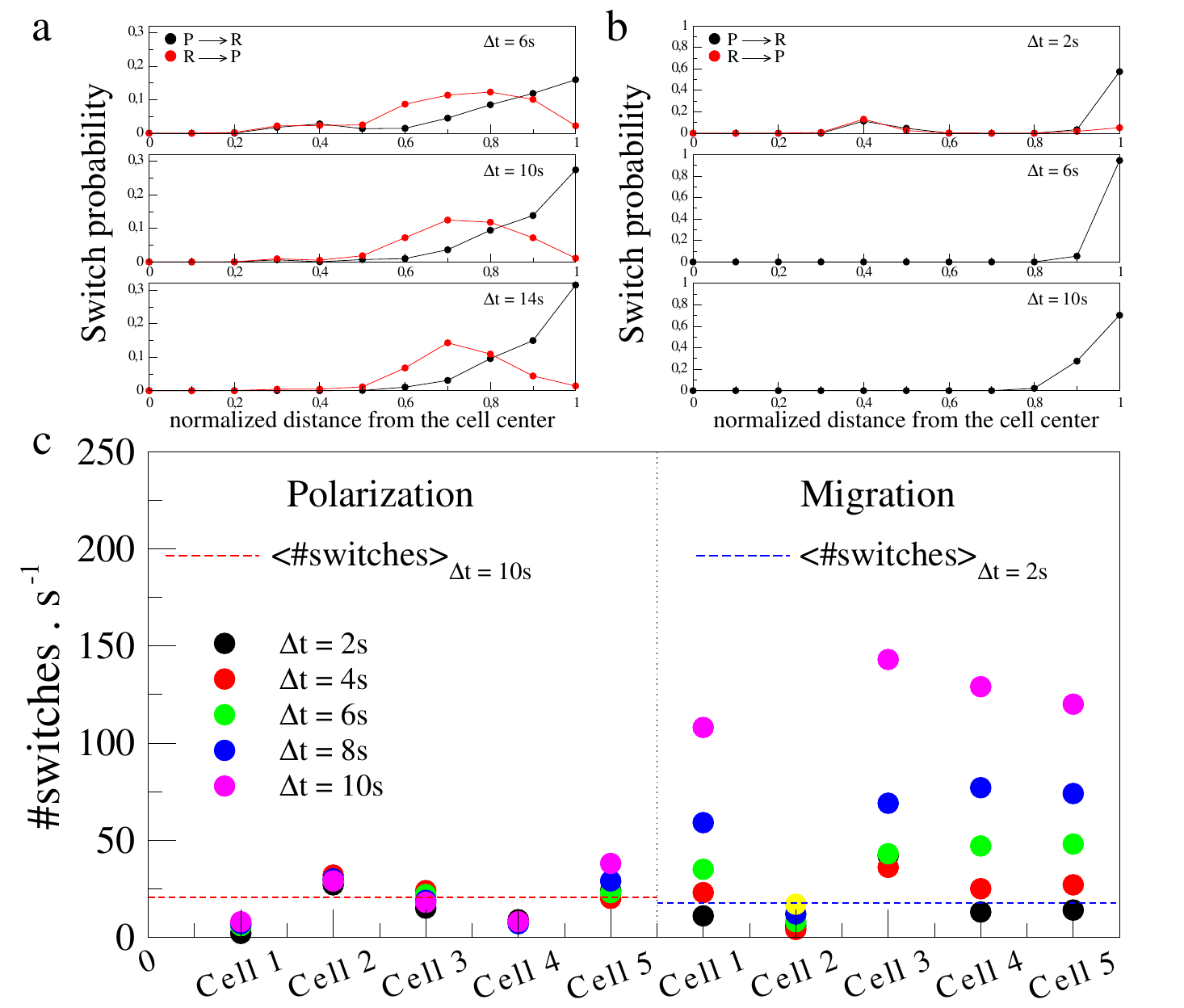}
\caption{ {\bf Switch probability distribution and switch frequency in time measured at different time intervals.} a) PR and RP switch probability as function of the normalized distance from the cell center measured at time interval of 6s, 10s and 14s (from top to bottom) during polarization. b) PR and RP switch probability as function of the normalized distance from the cell center measured at time interval 2s, 6s and 10s (from top to bottom) during migration. c) Switch frequency in time during polarization (left panel) and migration (right panel) measured at different time intervals. Red (resp. blue) dashed line marks the average number of switches per time unit during polarization (resp. migration) at a time interval of 10s (resp. 2s).}\label{figS1}
\end{figure*}
\newpage

\begin{figure*}[h!]
\centering \includegraphics[ clip=true,width=\textwidth]{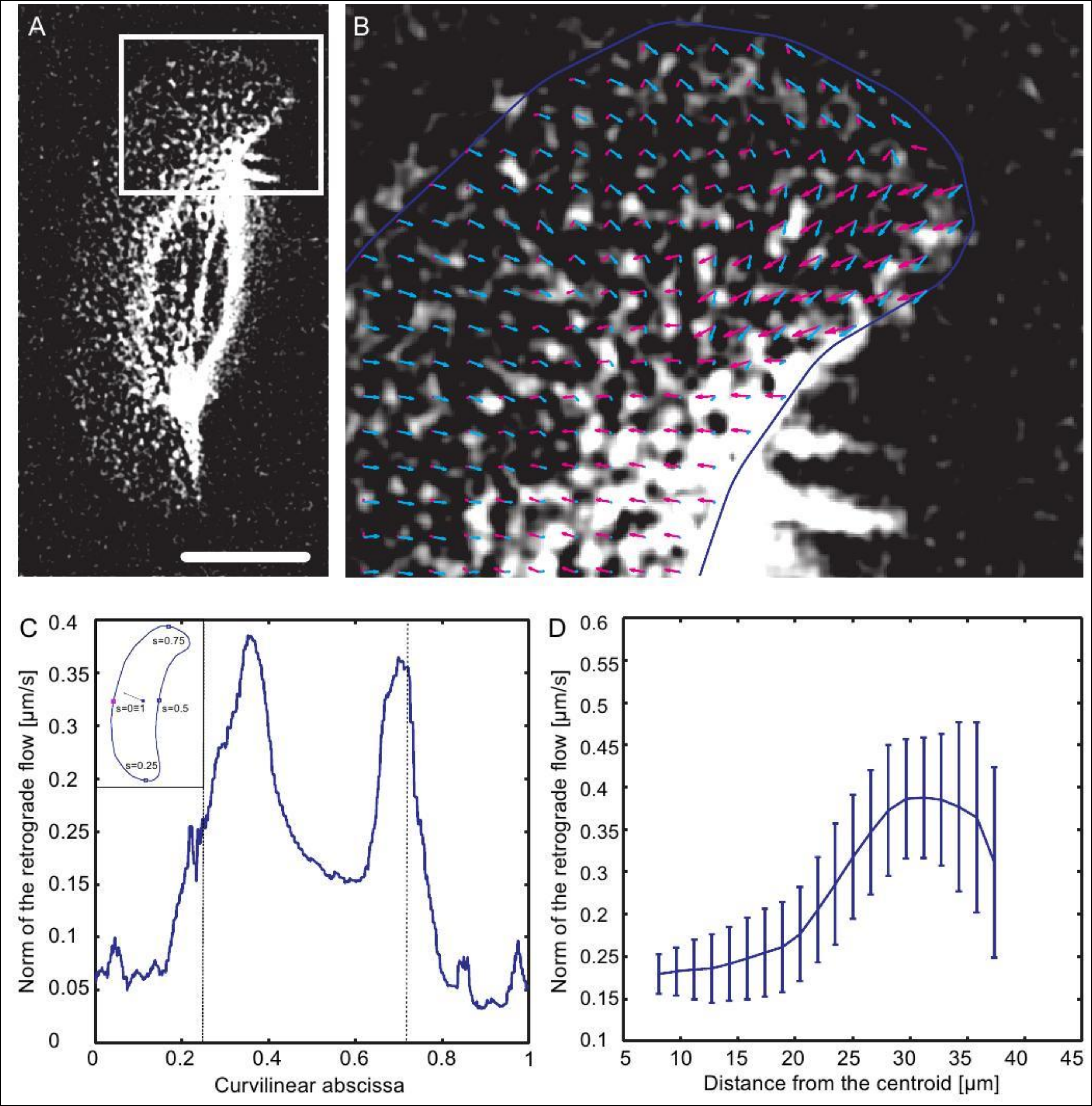}
\caption{ {\bf Retraction dynamics as visualized by fluorescence speckle microscopy of injected Alexa-phallodin.} (A) Fluorescence microscopy image of a crawling cell. Scale bar is 20 mm. (B) Close-up of the region enclosed by the white rectangle in (A) overlaid with the cell contour (blue, extracted manually), and the averaged cytoskeletal flow field over 28 s in the cell frame of reference (cyan) and in the laboratory frame of reference (magenta). (C) Actin flow velocity at the cell edge in function of the curvilinear abscissa averaged over 40 s. The inset shows the definition of the curvilinear abscissa along the cell edge with the centroid and the direction of motion indicated inside the contour. (D) Flow velocity in function of the distance from the centroid, averaged over 40 s. Data are obtained from points at the trailing edge with curvilinear abscissa between 0.25 and 0.71 (dashed lines in (C)), error bars show standard deviations.}\label{figS2}
\end{figure*}
\newpage

\begin{figure*}[h!]
\centering \includegraphics[width=0.875\textwidth]{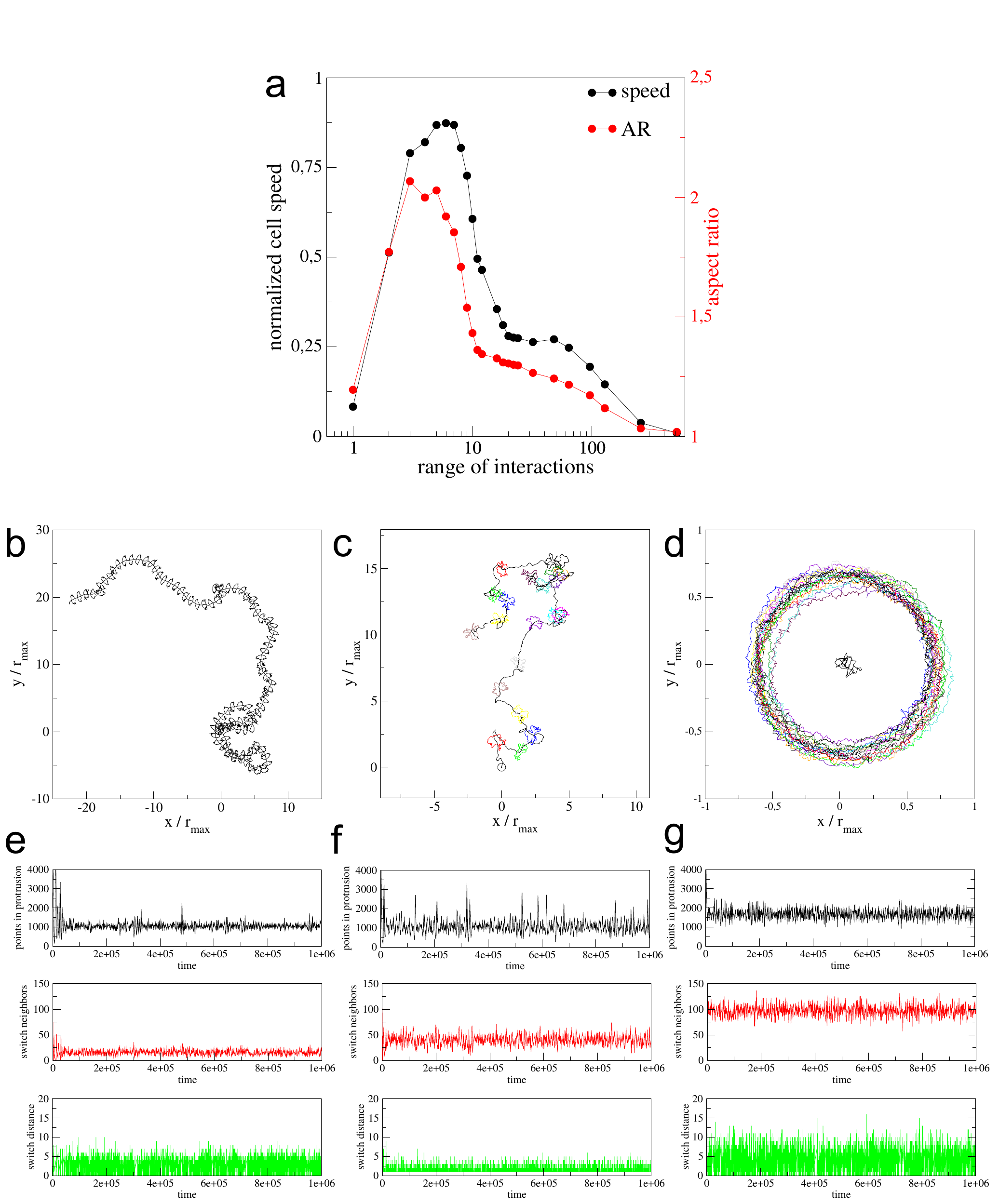}\vspace{-0.25cm}
\caption{{\bf Influence of the range of interaction N on the cell edge dynamics.}
(a)  Average cell velocity normalized by the velocity of protrusion Vp  (black curve), and the average cell outline aspect ratio (red curve) vs. range of interaction. For $2\le N \le 10$, the average cell speed is $\ge 0.5V_p$ (the system is considered to be in a motile state) with elongated shapes (AR $\le 1.5$). Simulation lasted for $10^6$ time steps and were repeated 10 times for each range of interaction. Typical trajectories and cell outlines for N=6 (b), N=16 (c) and N=512 (d). Oulines were separated by 10,000 time steps (b) and 50,000 time steps (color curves (c,d)). We observed the transition from persistent motion with stable elongated shape (b) to irregular motion with disordered shapes (c) and a completely immobile state with rounded shapes (d). Timeseries of the number of points in protrusion (black), PR switches due to neighbor interaction (red), PR switches due to distance threshold (green) for N=6(e), 16(f) and 512(g).}\label{figS3}
\end{figure*}
\newpage

\begin{figure*}[h!]
\centering \includegraphics[width=0.8\textwidth]{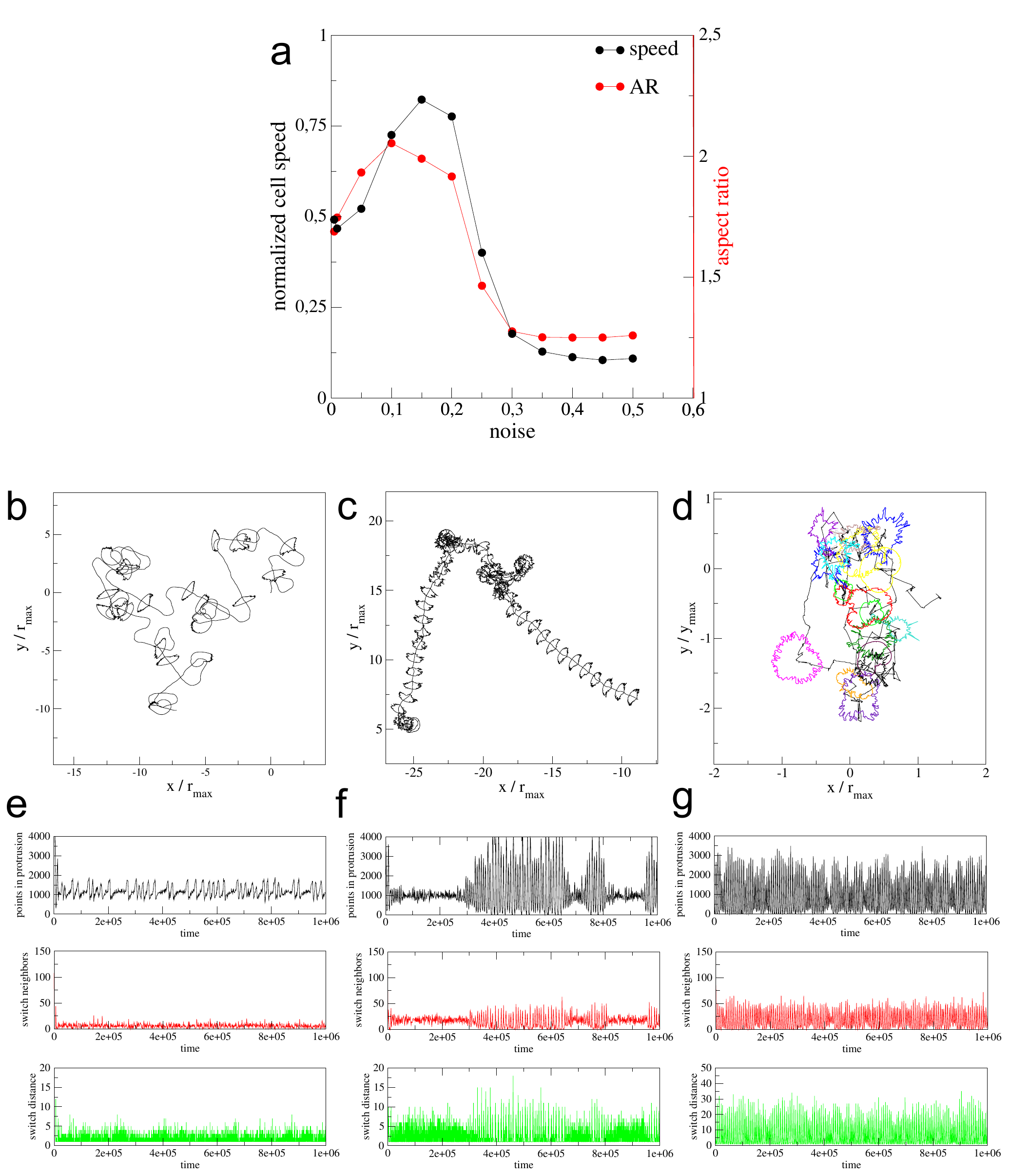}\vspace{-0.5cm}
\caption{{\bf Influence of the noise $\sigma$ on the cell edge dynamics.}
(a) Average cell velocity normalized by the velocity of protrusion $V_p$ (black curve), and the average cell outline aspect ratio (red curve). Simulations lasted $10^6$ time steps and were repeated 10 times for each value of $\sigma$. We observed an optimal noise value  $\sigma \sim 0.15 r_{max}$, characterised by a maximal average cell speed and a high aspect ration. For smaller noise values ($\sigma<0.1$) shape and motion were unstable. Due to the small noise on the distance threshold, regions of protrusion initially at the front of the cell propagated to the back, resulting in a change of the cell shape and a turn in the direction of the propagation of the protrusive region. For higher noise value ($\sigma = 0.25 r_{max}$), the system was in a metastable state alternating irregularly between periods of persistent migration and no motion. At larger noise values ($\sigma\ge 0.3 r_{max}$), no stable motion and persistent shape were observed. Typical trajectories and cell outlines for  $\sigma=0.05r_{max}$ (b),  $\sigma=0.25r_{max}$ (c),  $\sigma=0.5r_{max}$ (d). Oulines were separated by 10,000 time steps (c) and 50,000 time steps (b and d). Timeseries of the number of points in protrusion (black), PR switches due to neighbor interaction (red), PR switches due to distance threshold (green) for $\sigma=0.05r_{max}$ (e), $\sigma=0.25r_{max}$ (f) and $\sigma=0.5r_{max}$ (g).}\label{figS4}
\end{figure*}
\newpage

\begin{figure*}[h!]
\centering \includegraphics[width=0.9\textwidth]{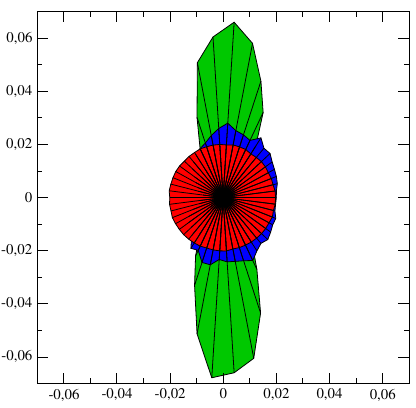}
\caption{{\bf Angular probability density distribution of protrusion-retraction interfaces.}
  Immobile cell (red, range of interaction N=512), non persistent moving cell (blue, range of interaction N=16) and persistent moving cell (green, range of interaction N=6). Coordinates of the interfaces are calculated in the frame of the cell, and angles are calculated with respect to the displacement of the cell between 1000 time steps. }
\label{figS5}
\end{figure*}
\newpage
\setcounter{figure}{0}
  \makeatletter 
\renewcommand{\figurename}{{\bf Supplementary Video}}
\makeatother

\begin{figure*}[h!]
\caption{{\bf Dynamics of fish epidermal keratocytes with cell bodies stalled by a pipette.} The sequence on the left shows an untreated cell, the middle sequence is in the presence of $50 \mu M$ blebbistatin, and the sequence on the right is in the medium diluted with $50\%$ water.}\label{vidpipette}
\end{figure*} 

\begin{figure*}[h!]
\caption{{\bf Keratocyte dynamics after cutting the lateral part of the cell with a glass microneedle.} Frame width $70 \mu m$, duration 3 min.}\label{vidcut}
\end{figure*}

\begin{figure*}[h!]
\caption{{\bf Overview of the process of polarization of fish keratocytes.} The cells were treated with 2.5 mM EDTA in 85$\%$ PBS to induce partial detachment and loss of polarity and then transferred back into the culture medium to allow for re-polarization. The video starts 1 min after the transfer to culture medium. Frame size 200 mm by 190 mm, duration 36 min.}\label{vidalicia}
\end{figure*}

\begin{figure*}[h!]
\caption{{\bf Different shapes and types of motile behavior of fish epidermal keratocytes.} Sequential movie clips represent typical persistently migrating cells with a single leading and a single trailing edge, a polarizing cell with several protruding and retracting regions consolidating into leading and trailing edges and cells with two and three leading edges propagating in circles around the cell perimeter.}\label{vidrotpol}
\end{figure*}

\begin{figure*}[h!]
\caption{{\bf Dynamics of a fish epidermal keratocyte with the cell body blocked by a pipette in the presence of $10\mu M$ nocodazole (to depolymerize microtubules).} The cell retracts and backs up from the obstacle similarly to untreated cells.}\label{pipettenoco}
\end{figure*}

\begin{figure*}[h!]
\caption{{\bf Computer simulation of cell-edge dynamics: Self-polarization and migration} Simulation based on the model of edge dynamics with protrusion/retraction switching probabilities depending on the distance from the cell center with the same set of parameters as specified in Methods.}\label{simulpolarization}
\end{figure*}

\begin{figure*}[h!]
\caption{{\bf Computer simulation of cell-edge dynamics: Comparison between distance threshold and area constraint.} The sequence on the left shows a simulation based on the model of edge dynamics with protrusion/retraction switching probabilities depending on the distance from the cell center. In the sequence on the right, switching probabilities depend on the cell area. Apart from these two different mechanisms of feedback control, the two simulations are run with the same set of parameters as specified in Materials and Methods. The outline with distance-dependent switching (protruding/retracting regions shown in green/blue, respectively) polarizes and exhibits persistent motion with a stable shape. The outline with area-dependent switching exhibits temporal separation of protruding and retracting zones (shown in red and black, respectively), but shows no stable shape, nor persistent motion.}\label{simulpolmig}
\end{figure*}

\begin{figure*}[h!]
\caption{{\bf Computer simulation of a keratocyte with three rotating leading edges.} The simulation is run with the parameters described in the caption of Fig.~5.}\label{simulrot}
\end{figure*}

\begin{figure*}[h!]
\caption{{\bf Computer simulation of keratocyte motion with the cell body blocked by a virtual pipette.} The gray line represents the “pipette” that the center of the cell cannot pass. Protruding/retracting zones are shown in black/red, respectively.}\label{simulpipette}
\end{figure*}

\begin{figure*}[h!]
\caption{{\bf Computer simulation after cutting the lateral part of a model keratocyte.} Edge points outside the cut line were removed, and new edge points were placed along the cut line with initially protruding states.}\label{simulcut}
\end{figure*}